\documentstyle[epsf,12pt,epsfig]{article}

\newcommand{\ls}{\mbox{$ l_{s} $}}

\newcommand{\lj}{\mbox{$ l_{11} $}}
\newcommand{\be}{\begin{equation}}
\newcommand{\br}{\begin{eqnarray}}
\newcommand{\ee}{\end{equation}}
\newcommand{\er}{\end{eqnarray}}

\begin{document}
\title{
\hfill\parbox{4cm}{\normalsize IMSc/98/10/51 \\hep-th/9810069}\\ 
\vspace{2cm}
The Hagedorn Transition, Deconfinement and the AdS/CFT Correspondence}
\author{ S. Kalyana Rama and B. Sathiapalan \\
{\em Institute of Mathematical Sciences}\\
{\em Taramani, 
Chennai 600113, INDIA }}

\vspace{5ex}
Email: krama, bala@imsc.ernet.in

\maketitle
\begin{abstract}
A connection between the Hagedorn transition in string theory and the
 deconfinement transition in (non-supersymmetric) 
Yang-Mills theory is made using the AdS/CFT correspondence. We modify
the model of zero temperature QCD proposed by Witten by compactifying
 an additional space-time coordinate with supersymmetry breaking 
boundary conditions thus introducing a finite temperature in the
 boundary theory.
 There is a Hagedorn-like transition associated with
winding modes around this coordinate, which signals a topology
changing phase
 transition to a new AdS/Schwarzschild blackhole where this coordinate
is the time coordinate. In the boundary gauge theory 
 time like Wilson loops acquire an
 expectation value above this temperature.

\end{abstract}
\newpage
\section{Introduction}

The question of whether the Hagedorn
temperature \cite{H}
is really a limiting temperature or whether it is to be interpreted as
a phase transition temperature is not fully understood.
 One motivation for interpreting
it as a phase transition comes from the belief that large-N Yang-Mills
theory has a string description, where the string coupling
constant, $g_s$, can be identified with $1/N$.  This theory is believed to have
a transition (for any $N$) from a low temperature confining phase
where the 
lightest
degrees of freedom are colour singlets, to a high temperature
deconfined
phase where the light degrees of freedom are gluons (and quarks). 
It is only natural to identify the Hagedorn temperature with the
temperature
of the deconfinement transition.
However the nature of the phase above that temperature is not fully
understood. In a recent work by one of us \cite{BS1}, using a matrix
model
 description
a picture of the phase transition was obtained that lends credence to
the identification of the Hagedorn transition as signalling 
something like a deconfinement transition. It was found that at 
high temperatures
D-0-branes tend to cluster and the uniform distribution that is
necessary in order to form a string is destroyed. 
The transition from uniform distribution to clustering is very similar
to that found in gauge theories at the deconfinement transition
temperature.
The expectation value of the time-like Wilson loop $<Tr e^{i\int A_0
dt}>$ vanishes when the eigenvalues of $\int A_0 dt$ are uniformly
distributed and becomes non-zero when the eigenvalues cluster.
Thus the high
temperature phase is a cluster of D-0-branes.

One can ask this question again in the light of some of the recent
developments that relate string theory in AdS backgrounds with
supersymmetric gauge theories [\cite{JM1}-\cite{GO}]. Here, even more
than in matrix theory, the connection between gauge theories and
string theory is very direct. Particularly useful is the fact that the
expectation value of a Wilson loop is given by summing over string
world sheets
that have this loop as their boundary \cite{JM2,RY}.

In \cite{EW2} a proposal was made to study non-supersymmetric
QCD in this approach. Several qualitative features expected in QCD
were 
demonstrated
such as the area law for Wilson loops and the existence of a mass gap.  
Following this proposal masses of the glueball states were computed
and reasonable agreement was found with lattice calculations
\cite{COOJ,MKJM,Z}. It
has also been pointed out that the AdS/CFT correspondence has been
tested in the regime where $g^2 N$ is large and therefore cannot be
said to describe continuum QCD but rather QCD on a very coarse
lattice. In this regime one cannot expect any kind of universality,
i.e. the details of the bare theory and the regularization become 
important. In particular one should not expect all physical quantities
to be some pure number (dependent only on $N$) times some appropriate
power of $\Lambda _{QCD}$. Presumably this explains the factor of
$g^2 N$ relating the string tension to glueball mass
\cite{GOl}. Furthermore
it is not clear that the continuum theory that one can get from this
is the usual QCD or QCD along with an infinite number of other
particles
coming from the Kaluza Klein states \cite{GO,ORT}. Ignoring these problems
for the moment
one can go ahead and address the issue of the connection between
the Hagedorn transition and deconfinememt. \footnote{A deconfinement transition
is expected in the lattice strong coupling theory also \cite{Svet}. So this is 
still a pertinent question.}
Various issues concerning the
Hagedorn transition have been discussed in recent papers \cite{CM,BKR}.

Our starting point is the proposal of \cite{EW2} for studying
non-supersymmetric QCD. Non-extremal M-5 branes in Euclidean $AdS_7 \times S_4$
are 
described
by the
following  metric: ($u,x^i$ are dimensionless. $\lj$ is the Planck
length in eleven dimensions.)

\[
{ds^2 \over (\lj)^2} = {u^2 \over {(\pi N )^{1/3}}}\sum _{i=1}^5
(dx^i)^2 \; + \; {u^2 \over {(\pi N )^{1/3}}}(1-{u_0^6 \over u^6})dx_0^2 
+ \,  4(\pi N)^{2/3}{1\over (1-{u_0^6 \over u^6})}{du^2\over u^2} 
\]
\be
 +(\pi N)
^{2/3} 
d\Omega _4 ^2
\ee
The time direction is compactified with a period $\frac{4\pi}{3}
\sqrt {\pi N \over
u_0^2}$
and supersymmetry is broken by the finite temperature boundary conditions.
One more direction is  compactified  with 
 supersymmetric boundary conditions.
 Thus one ends up with a limiting form of an  AdS Schwarzschild black hole or
 non extremal D-4 brane metric with one compact Euclidean time direction
to give a four (non compact Euclidean) 
dimensional 
boundary gauge
theory with massive fermions (supersymmetry is broken). This theory is
4-dimensional Euclidean QCD.
 By the rescaling ${u \over u_0} =  {\rho \over b}$
with $b= (\pi N )^{1/3}\lj $
 and 
$ {u_0 \over
\sqrt {\pi N} }x^i  =  z^i$ i=1,..4;  $x^5 = \frac{R_{11}}{\lj}\theta$ and
$\phi = \frac {3u_0}{2\sqrt {\pi N}} x_0$ ,
one ends up with the form ($R_{11}, b$ have dimensions of length and
so
$\rho $ has dimensions of length and all other coordinates are dimensionless.) 
\be \label {3}
ds^2 = \frac{4}{9}\rho ^2 (1- {b^6 \over \rho ^6}) d\phi ^2 + {4b^2 \over
 (1- {b^6 \over \rho ^6})} {d\rho ^2 \over \rho ^2} +
\rho ^2 \sum _{i=1}^{4}(dz^i)^2 + b^2d \Omega _4 ^2 + \frac{u_0 ^2
 R_{11}^2}{\pi N \lj ^2}\rho ^2 d\theta ^2
\ee
where  $\theta$, the ``eleventh'' coordinate and $\phi$ the
Euclidean time/temperature coordinate have periods $2\pi$.  
If we ignore $\theta$, (\ref{3})
represents a 10-dimensional space-time appropriate for string theory.
The dilaton field satisfies $e^{2\phi /3} = {R_{11} u\over \lj (\pi N)^{1/6}}$.
 The `string metric' where $\ls$ is treated as a constant can be
written as follows: ($w = g_s^{1/3}u^2 $ and $y^i = x^i g_s ^{1/3}$)
\footnote{Note that $g_s$ is defined as the asymptotic value of
$e^{\phi}$
at infinity where space is flat. This is $(\frac{R_{11}}{\lj})^{3/2}$.
The expression for the dilaton given
above
is only valid in the near horizon region where the metric is AdS. Thus
asymptotically $(\lj)^2 = g_s ^{2/3} (\ls)^2$, whereas in the AdS region where
we are working $(\lj)^2 = g_s ^{2/3}\frac{u}{(\pi N)^{1/6}} (\ls)^2$.}
\be \label{4}
{ds^2 \over \ls ^2}= 
{w^{3/2}\over \sqrt {g_s N\pi}}
(1-\frac{w_0^3}{w^3})dy_0 ^2+
{w^{3/2}\over \sqrt {g_s N\pi}}
\sum _{i=1}^4 (dy^i ) ^2 
+ \frac{\sqrt {g_s N\pi}}{ w^{3/2}  (1-\frac{w_0^3}{w^3})}dw^2
+\sqrt{g_s N\pi w} d\Omega _4 ^2
\ee

Note that while there is a temperature 
(=$\frac{3}{4\pi}\sqrt{{w_0\over g_sN}}$)
associated with the black hole
and we are doing finite temperature string theory, from the viewpoint
of the gauge theory we are still at zero temperature. The time
ccordinate
of the black hole metric is not one of the four coordinates 
$y^i, i=1,..4$ of the
gauge theory.
We, however, would like to study the gauge theory at finite temperature.
Accordingly we compactify one of the Euclidean directions of the gauge theory
(say $z_4=$`$z$')
and thus also of string theory, with {\em  supersymmetry  breaking}
boundary conditions. The radius $R_z$ of this circle is $\beta \over 2\pi$,
where
$\beta$ is the inverse temperature of the {\em gauge theory}. 
\footnote{We remind the reader 
that the string theory is already at finite temperature 
 identified
with the radius
$\frac{2}{3}\sqrt{g_sN\over w_0}$ 
of the compact Euclidean {\em time} coordinate $y_0$.}
>From the viewpoint of string theory we have a compactification of
Type II string theory with one compact direction `$z$' for
supersymmetry breaking (first studied in \cite{RR}) at finite temperature.
Thus the proposal of \cite{EW2} is modified in that
over and above finite temperature effects there is a supersymmetry
breaking due to boundary conditions in the $z$-direction. Thus the
zero-temperature string theory itself has broken supersymmetry.
 As shown in \cite{RR} this non supersymmetric
string has a tachyon in the sector with non zero winding in the  
$z$-direction when the radius is smaller than a critical value
of $O(\ls )$. 
This is similar to the tachyonic mode
found in the time-like direction in finite temperature string theory,
signalling the Hagedorn transition
\cite{BS2,IK}. In the (finite temperature) supersymmetric case one had 
to explain why this
tachyon is not projected out by the GSO projection. This was done in
\cite{AW}. For similar reasons the tachyon in the 
winding sector in the $z$-direction is not projected
out by the GSO projection.

Thus for a critical value of $\beta$ one expects a Hagedorn-like
phase transition. We would like to argue that this signals a
deconfinement transition in the {\em boundary gauge theory}. To make 
a connection with deconfinement one has to show that the expectation
value
of the time like Wilson loop (or Polyakov loop) is non-zero , i.e. 
\[
<P(C)> \equiv <Tr e^{i\int _c A_0 dt}> \, = \, 0  \; \; \; \beta > \beta _c 
\]
\be \label {2}
<P(C)> \equiv <Tr e^{i\int _c  A_0 dt}> \, \neq \, 0  \; \; \;  
\beta \leq \beta _c \\
\ee
One expects in continuum QCD that $T_c$ will be of the same order as
 the string tension. We will in fact see that the Hagedorn transition
 is of this order. \footnote{The actual $T_c$ will turn out to be 
 a little different - in fact 
it is of the order of the mass of the glueball. Again, as mentioned
above the fact that the glueball mass and the string tension are not
 the 
same must be an
 artifact
of the fact that we are doing a strong coupling calculation and are
far from the continuum limit.}

Let us remind ourselves why $P(C)$ is zero at low temperatures
\cite{EW2}. According to the prescription for calculating Wilson loops
one has to sum over surfaces of which this loop $C$ is the boundary.
In the following discussion, for convenience we will refer to $y_0$
as $t$ and $y_4$ as $z$.
 If we compactify the $z$-coordinate in the usual Kaluza-Klein manner,
 the $S^1$
around that direction is non-simply connected.  Thus this  $S^1$ is
not
the boundary of any two dimensonal surface that has the topology of a
disc.
 Hence $<P(C)>=0$. 
In order for $<P(C)>$ to become non-zero
at high temperatures,  $S^1$ must somehow become the
boundary
of some surface. The space time manifold must change topology for this
to happen. Now the time
coordinate of a 
non extremal D-brane  black hole is simply connected - if the
$S^1$ had been around this coordinate it would be the boundary of a
two
dimensional surface and $P(C)$ would be non-zero.
Thus if it so happens that the topology changes
so that $z$ becomes the Euclidean time coordinate
(and so $t$ becomes just another space coordinate)
 $P(C)$ will become non zero. 
It is plausible that such a tunneling can take place since in
Euclidean
space all the 10 coordinates are on equal footing and {\em a priori}
any of them can be analytically continued to a Minkowski time
coordinate.
In any case one must sum over all bulk manifolds whose boundary
is $S^1 \times S^1 \times \Sigma _3 $ spanned by $y^0 , y^4$ and
$y^i, i=1,2,3$ respectively \cite{EW1,EW2}.  Thus it is imperative
that
these topologies be considered in calculating the partition function.
One can compare the free energies of the two configurations
i.e. one, say $X_t$, where $t$ is time and and the other $X_z$
where $z$ is time, and see which configuration is dominant as
$R_z$ or equivalently the gauge theory temperature $=\frac{1}{2\pi
R_z}$
is varied. By
symmetry
it is clear that there will be an inverse transition temperature $\beta _c =
2\pi (R_z)_c$ at
which $X_z$ is preferred and furthermore (again by symmetry) 
that $\beta _c$ must be reached
when the two radii $R_t$ and $R_z$ are equal. We will be more 
explicit in the next section.

Thus we see that at a critical temperature there will be a phase
transition at which $P(C)$ acquires a non-zero vacuum expectation value. 
>From the string viewpoint it is a change of topology of space time.
This is the deconfinement transition for the boundary gauge theory.
 In fact we will see that this critical
temperature
will be lower than the ``Hagedorn'' temperature of the
$z$-drection. This is just as well. At the Hagedorn temperature, the
partition
function diverges and therefore the actual phase transition should
take place before that. This is also consistent with the results of
the weak coupling
string perturbation calculation of \cite{AW}.

\section{Calculation of the Deconfinement Temperature}

We will use the metric (\ref{3}) and assume that the boundary
gauge theory is located at some value of $\rho$ say $\rho _{\infty} >>
b$.
We will also take the boundary metric to be $ds^2 = \rho _{\infty}^2 
\sum _{i=1}^4 (dz^i)^2$.  Thus the value  $\rho _{\infty}$ can be
thought of as a scale in the renormalization group sense. Using
the usual arguments the string tension of the boundary gauge theory
is $\sigma = \frac{b^2}{\rho _{\infty}^2 \ls ^2}$.  We must remember
that  $\lj = e^{\phi /3} \ls$ and hence
in the metric of (\ref{3}) $\ls$ is $u$ dependent. If we go to the string
metric
we find $\sigma = (\frac{w_0}{w_\infty})^{3/2}\frac{1}{\ls ^2}$, where
$w_{\infty}$ is the value of $w$ of the boundary gauge theory (see
(\ref{4})).
If in (\ref{3}) we compactify say $z^4 = z$ with a radius 
$R_z$ with supersymmetry
breaking
boundary conditions, we get a boundary gauge theory with inverse
temperature $\beta = \rho _{\infty} 2\pi R_z$. As we reduce $R_z$
the winding mode becomes tachyonic first near $\rho = b$. this happens
when $2\pi R_z b \approx \ls$. Thus the temperature at the boundary
when this Hagedorn-like transition takes place is $\approx \frac{b}{
\rho _{\infty}  \ls } \approx \sqrt \sigma $.  This is of course when
the deconfinement transition is also expected to take place. This
supports the idea that a Hagedorn transition in the bulk signifies a
deconfinemnt transition in the boundary.

The value of $2\pi R_t$, the circumference  of the time coordinate at 
$w_{\infty}$,
is ${4\pi \over 3}\frac{w_{\infty}^{3/4}}{(g_s N)^{1/4}}\sqrt{\frac{g_sN}{w_0}}\ls ={4\pi \over 3}
(g_sN)^{1/4}\frac{w_{\infty}^{3/4}}{w_{0}^{1/2}}\ls $.
Thus the ratio  $R_z \over R_t$ when the Hagedorn transition takes
place
is $\frac{1}{(w_0 g_s N )^{1/4}} $ which is expected to be less than
one
as $g_s N $ is much greater than one, unless $w_0$ is very small.  In
fact, as $R_z$ is reduced (thus raising the gauge theory temperature),
 before the Hagedorn transition is reached at $R_z < R_t$, 
when $R_z$ becomes equal to $ R_t$ we expect a phase transition to a 
spacetime where $z$
is the time coordinate. This can be easily seen by comparing the
corresponding expressions for the free energy \cite{BKR}. 
Let $F(R_z , R_t)$ be
the
expression for the free energy when $t$ is the time coordinate
(we do not need the exact expression here).
  When $z$ becomes the time coordinate we get $F(R_t, R_z )$  (by
symmetry) - the same function with $R_z$ and $R_t$ interchanged.  We
know that when $R_z$ is infinite, the preferred configuration is the
one where 
$t$ is the time coordinate. Therefore it must be that
 $F(R_z, R_t) < F(R_t ,R_z )$ when $R_z$ is very large.
Clearly
the transition to the other regime, which we have seen makes $< P(C)
>$
 non zero,
  occurs when $F(R_z, R_t) = F(R_t ,R_z )$ which means $R_t = R_z$.

Thus unless $w_0$ is very small we expect the deconfinement transition
to take place before the Hagedorn.  The temperature at which deconfinement
happens (i.e. when $R_z =R_t$), 
is $\frac{1}{2\pi R_t}\approx O(M_{glueball})$\cite{COOJ,MKJM,Z}.  As mentioned
earlier,
the fact that this is not equal to the  string tension must be an artifact
of the strong coupling limit. Presumably if we knew how to go to weak
coupling these two scales would converge.

To establish that the entropy does jump by an amount proportional to 
$N^2$, \cite{Th,EW1}, we need to look at the expression for $F$.  
Actually it is
trivially proportional to $N^2$ always because of the presence
of Newton's constant. But, as explained in \cite{EW1} the part that 
is linearly dependent on $\beta$ is just the zero temperature
contribution. 
What we have to look at is the coefficient of the piece that has a 
non trivial $\beta$ dependence. Another way of saying the same thing
is that the thermodynamic free energy is $-ln Z \over \beta$ whereas
we have been referring to $-ln Z$ as the free energy. After dividing
by $\beta$, the piece that is temperature dependent should be
proportional
to $N^2$ above the deconfinement transition, whereas below the
transition
this piece will be of $O(1)$. 
Remembering that $R_z$ is $ \beta \over 2\pi$ we can easily
see that $F(R_t, {\beta \over 2\pi})$ must have the form (in the
supergravity
approximation, when the fields or metric have no dependence on $\beta$) 
$\beta N^2 f(R_t )$ which is linear in $\beta$ and does not contain
any non linear $\beta$ dependence. Thus after dividing by $\beta$
there is no temperature dependent contribution in this approximation.
 This means in particular that there are no
finite temperature effects that are proportional to $N^2$. 
( We remind the reader that by `temperature' we mean the
temperature of the boundary four dimensional gauge theory and {\em not}
the temperature of the AdS-Schwarzschild black hole.) 
The function $F({\beta \over 2\pi}, R_t )$ on the other hand
is of the form $2\pi R_t N^2 f({\beta \over 2\pi})$  which definitely
has a non trivial $\beta$ dependence proportional to $N^2$.
 We do not need the exact expression for $f$ to make these arguments.
We only need to realize that it cannot be linear in its argument and
therefore
$f({\beta \over 2\pi})\over \beta $ is still temperature
dependent. This is due to the non-trivial $\beta$-dependence of the metric. 
The reader can verify this using the exact expression given
in\cite{BKR} 
for the case at hand.
Thus this transition has all the characteristics expected of
a deconfinement transition in gauge theories.

\section{Conclusion}

To summarize we have shown that one can relate the deconfinement
transition
of finite temperature QCD to a Hagedorn-like transition in the bulk
string
theory. The Hagedorn transition involves condensation of vortices
around
a compact coordinate `$z$' with supersymmetry breaking boundary conditions.
This signals a phase transition where the topology of
spacetime
changes. The new topology is one where $z$ becomes the time coordinate
and the old time coordinate becomes a space coordinate. The
expectation value of the Polyakov loop becomes non zero and therefore
this transition  can be identified with deconfinement.

When we combine the above arguments with the results of \cite{BS1} the
following picture of the deconfinement transition emerges: 
As one heats up the gauge theory 
 there is a tendency for the D-0-branes making up the membrane
(we are using the matrix model \cite{BFSS} language here) that
becomes a string when wound around a compact direction, 
to cluster. This is because supersymmetry is broken by the boundary
conditions on the $z$-direction. This is 
 explained in \cite{BS1}. One can think of this as a tendency of the string to
shrink to zero size. This makes the $z$-coordinate pinch off to a
point so that the string can shrink to zero size at the tip near the
region of small $u$ and disintegrate into a bunch of D-0-branes. Thus
near (inside?)
the horizon we have a cluster of D-0-branes and no string. 
This disintegration happens at the small $u$ end first because the
temperature
increases as you move towards the horizon and is highest at the
horizon.
 Just when the $z$-coordinate
pinches off and $<P(C)>$ becomes non-zero, the
$t$-coordinate
opens up and becomes an ordinary non-simply-connected circle.

\end{document}